\documentclass[useAMS,usenatbib]{mn2e}

\title[Nulls and subpulse drift]{Instabilities nulls and subpulse drift in radio 
pulsars}
\author[P. B. Jones]{P. B. Jones\thanks{E-mail:
p.jones1@physics.ox.ac.uk}\\ 
Department of Physics, University of Oxford, Denys Wilkinson Building,\\
Keble Road, Oxford OX1 3RH}

\begin{document}

\date{}

\pagerange{\pageref{firstpage}--\pageref{lastpage}} \pubyear{}

\maketitle

\label{firstpage}

\begin{abstract}

This paper continues a previous study of neutron stars with positive polar-cap 
corotational charge density in which free emission of ions maintains the surface 
electric-field boundary condition ${\bf E}\cdot{\bf B} = 0$.  The composition of 
the accelerated plasma on any subset of open magnetic flux-lines above the polar 
cap alternates between two states; either protons or positrons and ions, of 
which the proton state cannot support electron-positron pair creation at higher 
altitudes.  The two states coexist at any instant of time above different moving 
elements of area on the polar cap and provide a physically consistent basis for 
a description of pulse nulls and subpulse drift.  In the latter case, it is 
shown that the band separation $P_{3}$ is determined not by the ${\bf 
E}\times{\bf B}$ drift velocity, as is generally assumed, but by the diffusion 
time for protons produced in reverse-electron showers to reach the region of the 
atmosphere from which they are accelerated.  An initial comparison is made with 
the survey of subpulse drift published by Weltevrede et al.

\end{abstract}

\begin{keywords}

pulsars: general -stars: neutron - plasma - instablities

\end{keywords}

\section{Introduction}

It is now generally assumed that electron-positron pair creation  immediately 
above the magnetic poles is essential for coherent radio emission in pulsars. 
Muslimov \& Tsygan (1992) have made an important contribution in their 
recognition of the significance of the Lense-Thirring effect in the generation 
of adequately strong electric fields (see also the paper of Muslimov \& Harding 
1997). It is remarkable, and must be almost unique, that a general-relativistic 
effect changes the order of magnitude of a field component in a system that can 
be described, ostensibly, perfectly well in Euclidean space as in the paper of 
Goldreich \& Julian (1969).

At least initially, finding the acceleration field ${\bf E}_{\parallel}$ which 
is the component locally parallel with the magnetic flux density ${\bf B}$, can 
be regarded as a problem in electrostatics within a volume in the corotating 
frame of reference having well-defined boundary conditions. 
This is defined by the neutron-star surface and the surface separating open from 
closed magnetic flux lines and is assumed to have the electric potential 
boundary condition $\Phi = 0$. It is possible that this surface is actually a 
current sheet of some complexity (see, for example, Arons 2010), but we shall 
assume that, near the neutron-star surface, the time-dependence 
 of ${\bf E}_{\parallel}$ within it is not sufficiently rapid to hinder the 
charge movement necessary to maintain the boundary condition.  Thus
there are three possible cases depending, firstly, on the relation between 
rotation spin ${\bf \Omega}$ and ${\bf B}$ and hence the sign of the polar-cap 
Goldreich-Julian charge density $\rho_{0}$, and secondly on the neutron-star 
surface value of ${\bf E}_{\parallel}$.  They are: (i) ${\bf \Omega}\cdot{\bf B} 
> 0$, $\rho_{0} < 0$, ${\bf E}_{\parallel} = 0$ with electron acceleration; (ii) 
${\bf \Omega}\cdot{\bf B} < 0$, $\rho_{0} > 0$, ${\bf E}_{\parallel} = 0$ with 
positron and positive baryon acceleration or (iii) ${\bf \Omega}\cdot{\bf B} < 
0$, $\rho_{0} > 0$ and ${\bf E}_{\parallel} \neq 0$.
Throughout this paper, for brevity, these conditions will be referred to as 
cases (i), (ii) or (iii).  There is no further possible case because free 
emission of electrons is always possible at neutron-star surface temperatures.
A previous paper (Jones 2010a, hereafter Paper I) reported some preliminary work 
on the relation between these sets of boundary conditions and observed 
phenomena, in particular, the existence of pulse nulls in a significant fraction 
of radio pulsars.  There seems to be no reason why neutron stars satisfying (i), 
(ii) and even exceptionally (iii) should not exist and it is the purpose of the 
present paper to continue the attempt to see if features of the predicted plasma 
acceleration in these cases correlate with observed properties of subsets of the 
pulsar population.

There can be little doubt that in this context nulls and subpulse drift are
important phenomena. Although there must be reservations about assigning undue 
weight to the properties of a single neutron star, nulls in the isolated pulsar 
PSR B1931+24 are informative. Kramer et al (2006) found that spin-down in the 
on-state is approximately twice as fast as in the off-state. It is difficult to 
see, in a pulsar of this age ($> 10^{6}$ yr), how the geometrical shape of the 
acceleration volume or the magnitude of the current density ${\bf J}$ within it 
can change by a factor of this order in less than the rotation period $P$. 
Time-varying fields near the light cylinder can further accelerate 
ultra-relativistic particles of both signs and therefore the most obvious 
explanation for the change in spin-down torque is that the particle and field 
components of the magnetospheric momentum density
and stress tensor
near the light cylinder differ between on and off-states. A cessation of pair 
creation during nulls would greatly change the charged particle number density 
even though ${\bf J}$ remains essentially unchanged. In this connection, it is 
interesting that a recent paper by Lyne et al (2010) suggests quasi-periodic 
switching between two different spin-down rates as the origin of the peculiar 
timing residuals seen in many pulsars.

The presence of nulls might be thought of as evidence that a pulsar is in the 
final phase of evolution prior to complete cessation of radio emission.
However, we suggest that this view is not consistent with a specific feature  of 
the 63 nulling pulsars listed in Tables 1 and 2 of the paper by Wang, Manchester 
\& Johnston (2007).  The maximum potential difference $\Phi _{max}$ available 
for acceleration at the polar cap is proportional to $B_{d}P^{-2}$, in which 
$B_{d}$ is the effective dipole field inferred from the spin-down rate.  It is 
approximately independent of the ratio of $B_{d}$ to the actual polar-cap field 
$B$.
The distribution of this quantity for all radio pulsars in the ATNF catalogue 
(Manchester et al 2005) has a relatively sharp cut-off at $2.2\times 10^{11}$ G 
s$^{-2}$, equivalent to the existence of a well-defined death line in the 
distribution of $B_{d}$ versus $P$.
It might be expected that nulls should be seen only as a pulsar rotation slows 
so that it approaches this value. But the form of the distribution of 
$B_{d}P^{-2}$ for the nulling pulsars listed by Wang et al is broadly 
indistinguishable from that of the whole ATNF catalogue and instead is 
consistent with nulls being a long-term property of a certain sub-set of radio 
pulsars.  It is, of course just possible that the observed sharp cut-off is 
itself a statistical fluctuation and that individual pulsar deaths actually 
occur at values of $B_{d}P^{-2}$ throughout the whole distribution because some 
other variable is involved, such as flux-line curvature.  But we shall discount 
this possibility and in the present paper adopt the view that pulsars with 
boundary condition (ii) and, exceptionally, (iii) form the sub-set that null.

A recent large-scale survey by Weltevrede, Edwards \& Stappers (2006) has 
revealed that subpulse drift is a common phenomenon.  Under the assumption that 
electron-positron pairs are the source of the coherent radio emission, it 
implies that compact zones of pair creation exist and move in an organized way 
within the open magnetosphere area of the polar cap. This motion has been 
assumed to be an
${\bf E}\times{\bf B}$ drift velocity under the case (iii) surface electric 
field boundary condition, following the original paper of Ruderman \& Sutherland 
(1975).  There have been many later papers on this subject and we refer to Gil, 
Melikidze \& Geppert (2003) for recent developments which have sought to refine 
the calculation of the drift velocity.  The problem is that the case (iii) 
boundary condition is implausible as a common property of the pulsar population 
and that cases (i) and (ii) have not hitherto provided any immediate and obvious 
basis for the formation of organized subpulses. It was shown in Paper I that 
short time-scale instability in the composition of accelerated plasma exists in 
case (ii). In this paper, it is shown to be a plausible mechanism for subpulse 
formation and drift.

Under the assumption of an actual dipole field, pair creation is possible, in 
principle, in all observed pulsars through the inverse Compton scattering (ICS) 
of polar-cap photons by accelerated electrons or positrons (Hibschman \& Arons 
2001, also
see Fig. 1 of Harding \& Muslimov 2002), but curvature radiation (CR)  can 
produce pairs only in those with high values of $B_{d}P^{-2}$. However, there is 
a problem, noted by Harding \& Muslimov, in the formation of a dense pair plasma 
because the number of ICS pairs formed per primary electron or positron 
accelerated can be smaller than unity (see also Fig. 8 of Hibschman \& Arons). 
They suggest that the pair density required for coherent radio emission may be 
far smaller than previously thought.  Although higher multipole field components 
may be present in most pulsars, so increasing flux-line curvature, the existence 
of this problem must remain a matter of concern perhaps even leading to doubts 
about the role of pair plasma as the source of coherent radio emission.  

The existence of solutions of the electrostatic problem under boundary 
conditions (ii) or (iii) is no more than a preliminary because we are concerned 
in this paper with the presence of instabilities which arise from the reverse 
electron flow at the polar cap. It might be questioned whether or not the charge 
density on the surface separating open from closed magnetic flux-lines needed 
for the condition ${\Phi} = 0$ can be maintained in the presence of instability.  
But we shall find that, even at short time-scales, instability principally 
changes the nature of the particles accelerated rather than the current density 
${\bf J}$ and the acceleration field.  The instabilities considered here are 
not, of course, a feature of case (i) in which electrons are the primary 
component of ${\bf J}$.  Establishing instability in this case is a purely 
electromagnetic problem which has been considered by Levinson et al (2005), 
Melrose \& Luo (2009) and Reville \& Kirk (2010). Non-stationary flow in case 
(iii) has been investigated by Timokhin (2010).

The properties of the condensed matter surface are naturally important in cases 
(ii) and (iii), in particular, the production of protons by the reverse flux of 
electrons described previously (Paper I; see also Jones 1981).  This is a 
characteristic of electromagnetic showers that is usually of little practical 
significance.  The form of the shower depends principally on electron-photon 
interactions but shower photons also interact directly with nuclei.  The 
cross-section is a maximum with the excitation of the giant dipole resonance 
(GDR), a broad collective state, at a photon momentum $k\approx 40$ mc.  
(Electroproduction cross-sections are smaller by a factor of the order of the 
fine-structure constant and can be neglected.)  Photon track length per unit 
interval of $k$ in a high-energy shower is $\propto k^{-2}$ so that to a good 
approximation,
photoproduction of baryons can be assumed to occur entirely through GDR 
formation.  Known cross-sections and electromagnetic shower theory allow the 
calculation of production rates, and we define $W_{p}$ as the number of protons 
produced per unit primary shower energy.  Protons are initially of the order of 
a few MeV but are very rapidly moderated to thermal energies without further 
strong interaction, and then diffuse to the surface with a time delay that is of 
prime significance for the stability of plasma acceleration.
For further details we refer to Paper I, also to Jones (2010b) for the 
cross-sections at high values of $B$ for processes of second order in 
electron-photon coupling.  In the early stages of acceleration, 
partially-ionized atoms interact with the polar-cap blackbody radiation field 
and this is the source of the reverse-electron flux that is considered here.  It 
has proved convenient to combine rates for this process with values of $W_{p}$ 
by defining, for a particular pulsar, the parameter $K$ which is equal to the 
number of protons produced per unit nuclear charge accelerated.

Instabilities in solutions of the time-independent electrostatic problem 
referred to above exist as a consequence of proton formation and we propose that 
they are the basis for the nulling phenomenon and for subpulse formation and 
drift.
The complexity that arises is unfortunate because it limits what can be achieved 
in terms of a physical theory of the acceleration process.  Thus we shall be 
able to expose the general properties attaching to cases (ii) and (iii) but are 
not always able to give quantitative predictions.
We shall show in Section 2 that general consideration of polar-cap parameters 
rule out the possibility that the relatively numerous subsets of pulsars that 
exhibit nulls and subpulse drift belong to case (iii). Therefore, sections 3 and 
4 of this paper are restricted to further consideration of 
case (ii) for which short time-scale instability in plasma acceleration has been 
described previously in Paper I. Its properties are summarized in Section 3.
The previous treatment of medium time-scales was, however, inadequate and 
contained an error.  The appropriate analysis is given here in Section 4.  The 
relations between these instabilities and the observed properties of nulls and 
of subpulse drift are described in Section 5 and, in particular, the relation 
beween the subpulse band separation periodicity $P_{3}$ and proton diffusion 
time is given.

\section[]{Polar-cap parameters}

Many of the polar-cap properties and parameters that will be required for the 
arguments of Sections 3-5 have been given previously in Paper I. The properties 
of the polar cap atmosphere are of particular importance and will be further 
considered in Section 3. But it will be convenient to summarize the remainder 
here with some additions.

The most basic parameter is the ion number density $N$ of the condensed matter 
at zero pressure. The magnetic dipole fields $B_{d}$ inferred from pulsar 
spin-down rates vary over about five orders of magnitude, but the median value 
for the 63 nulling pulsars listed by Wang et al is $2.8\times 10^{12}$ G, which 
is significantly larger than for the whole ATNF catalogue. It is also possible 
that the actual polar-cap field $B$ is larger than $B_{d}$.
For these reasons, we adopt the expression,
\begin{eqnarray}
N = 2.6 \times  10^{26}Z^{-0.7}B^{1.2}_{12}\hspace{5mm} {\rm cm}^{-3},
\end{eqnarray}
fitted to values computed by Medin \& Lai (2006), primarily for $B_{12} > 10$, 
where $B_{12}$ is the magnetic flux density in units of $10^{12}$ G and $Z$ is 
the atomic number.  The convenient unit of depth below the surface at $z = 0$ is 
the radiation length,
\begin{eqnarray}
l_{r} = 1.66Z^{-1.3}B^{-1.2}_{12}\left(\ln\left(12Z^{1/2}B^{-1/2}_{12}
\right)\right)^{-1}\hspace{5mm}{\rm cm}
\end{eqnarray}
defined here in terms of the zero-field Bethe-Heitler bremsstrahlung 
cross-section with screening factor modified for the neutron-star surface 
density (see Paper I).

The critical temperature above which the ion thermal emission rate is high 
enough to maintain the case (ii) boundary condition is related to the cohesive 
energy $E_{c}$ by $k_{B}T_{c} \approx 0.025E_{c}$ (see the discussion of 
atmospheric properties in Section 3). Cohesive energies have been calculated by 
Medin \& Lai as functions of $B$. For $Z = 26$ and $B_{12} = 10$, their value is 
in good agreement with that obtained by Jones (1985) using a different 
representation of the three-dimensional condensed matter state.
In the interval $10 < B_{12} < 100$, their values can be fitted by the 
expressions $E_{c} = 0.016B^{1.2}_{12}$ keV for $Z = 6$ and $E_{c} = 
0.16B^{0.7}_{12}$ keV for $Z = 26$ giving,
\begin{eqnarray}
T_{c} & = & 4.6\times 10^{3}B^{1.2}_{12}\hspace{5mm}{\rm K}, 
\hspace{1cm}Z=6\nonumber  \\
  & = & 4.6\times 10^{4}B^{0.7}_{12}\hspace{5mm}{\rm K},\hspace{1cm}Z=26.
\end{eqnarray}
These are to be seen in relation to other temperatures that are significant.
Paper I contained a calculation of the rate of proton formation in the 
electromagnetic showers formed by reverse electrons incident on the polar cap. 
From the energy flux needed to produce a proton current density $J^{p} = 
\rho_{0}c$, we can infer a maximum steady-state polar-cap temperature,
\begin{eqnarray}
\bar{T} = \left(T^{4}_{res} + \frac{(-B\cos\psi)(1 - \kappa)}
{P\sigma eW_{p}}\right)^{1/4}\hspace{5mm}{\rm K}.
\end{eqnarray}
In this expression, $W_{p}$ is the number of protons produced per unit primary 
shower energy.  We can approximate it initially by $W^{BH}_{p}$ which was 
obtained in Paper I by using the zero-field Bethe-Heitler pair creation 
cross-section with screening modified for the condensed matter density at the 
neutron-star surface. Its values, given in Table 1 of that paper, can be 
summarized conveniently in the intervals $B_{12} = 10^{1}-10^{2}$ and $Z = 
10-26$ by the expression 
\begin{eqnarray}
W^{BH}_{p} = 3.9\times 10^{-4}\langle Z\rangle^{-0.6}_{sm} 
B^{0.2}_{12}\hspace{5mm}(mc^{2})^{-1},
\end{eqnarray}
in which the nuclear charge is the average in the region of the shower maximum.
The angle $\psi$ is that between ${\bf \Omega}$ and ${\bf B}$.  The 
general-relativistic correction contained in the surface value of the Goldreich- 
Julian charge density $\rho_{0}$ is $\kappa$ (see Muslimov \& Harding 1997), and 
$\sigma$ is Stefan's constant.  Equation (4) also contains
a further temperature, $T_{res}$, which is the polar-cap temperature in the 
absence of any reverse electron or photon energy flux (approximately  the 
observed whole-surface temperature corrected to the local proper frame).  The 
presence of $T_{res}$ assumes that there is a constant heat flow to the surface 
driven by the very much higher temperature of the inner crust.  With $T_{res} = 
0$ and $\kappa = 0.15$, we find,
\begin{eqnarray}
T_{max} = 5.1\times 10^{5}\langle 
Z\rangle^{0.15}_{sm}B^{0.2}_{12}\left(\frac{-\cos\psi}{P}\right)^{1/4}
\hspace{5mm}{\rm K}
\end{eqnarray}
Equating this with $T_{c}$ allows us to estimate the minimum polar-cap magnetic 
flux densities necessary to sustain the case (iii) boundary condition ${\bf 
E}_{\parallel} \neq 0$.  These are,
\begin{eqnarray}
B_{12} & = & 181\left(\frac{-\cos\psi}{P}\right)^{1/4}\hspace{1cm}Z = 6\nonumber 
\\
       & = & 327\left(\frac{-\cos\psi}{P}\right)^{1/2}\hspace{1cm}Z = 26.
\end{eqnarray}
Comparison with the median value of $B_{d12} = 2.8$ for the 63 nulling pulsars 
listed by Wang et al shows that case (iii) can be widely realized only with 
implausibly large deviations from a central dipole field.

\begin{table}
\caption{The table gives values of $W_{p}$ for high magnetic flux densities $B 
\geq B_{c}$ and nuclear charge $Z$. The effective total cross-sections 
$\sigma^{P}$ for pair creation in the GDR photon energy band are for transverse 
momenta below the lowest magnetic conversion threshold at $k_{\perp} = 2$ mc and 
are in units of barns.  They
are estimates obtained from Table 3 of Jones (2010b) by averaging over photon 
polarization and over photon transverse momenta $k_{\perp} = 1.0$ and $1.5$ mc, 
and are equivalent to a mean free path for Coulomb pair creation lengthened by a 
factor $\eta_{p}$ compared with that for the modified zero-field Bethe-Heitler 
cross-section.  The Landau-Pomeranchuk-Migdal effect is not significant in the 
GDR photon energy band}

\begin{tabular}{@{}rcccr@{}}
\hline
 $Z$  &  $BB^{-1}_{c}$  &  $\sigma^{P}$  &  $\eta_{p}$  &  $W_{p}$  \\
      &                 &     bn         &        &  $(mc^{2})^{-1}$ \\
\hline
  10  &  1  &  0.27  &  1.14  &  $2.3\times 10^{-4}$   \\
      &  3  &  0.071  &  2.95  & $3.4\times 10^{-4}$   \\
	  & 10  &  0.017 &  6.3   &  $3.6\times 10^{-4}$   \\
  18  &  1  &  0.89  &  1.35  &  $1.8\times 10^{-4}$   \\
      &  3  &  0.23  &  3.8   &  $3.2\times 10^{-4}$   \\
	  & 10  &  0.055 &  9.7   &  $3.8\times 10^{-4}$   \\
  26  &  1  &  1.86  &  1.45  &  $1.3\times 10^{-4}$   \\
      &  3  &  0.48  &  4.2   &  $2.7\times 10^{-4}$   \\
	  & 10  &  0.11  &  11.3  &  $3.4\times 10^{-4}$   \\
\hline
\end{tabular}
\end{table}

However, the screening-modified zero-field Bethe-Heitler pair creation 
cross-section is not obviously valid at magnetic flux densities of the order of 
the critical field $B_{c} = 4.41\times10^{13}$ G.  Thus we have been obliged to 
calculate the second-order bremsstrahlung and pair creation cross-sections using 
Landau function solutions of the Dirac equation.
The photoproduction of protons by giant-dipole resonance (GDR) decay is 
determined by the total photon track length in the GDR band, centred on a 
momentum $k = 40$ mc,
which occurs almost entirely in the late stages of shower development. The track 
length at these energies is limited by Coulomb and magnetic
pair creation, also by Compton scattering the effect of which is almost always 
to scatter the photon so that its transverse momentum component $k_{\perp}$ 
(perpendicular to the field) exceeds the threshold for magnetic conversion to 
electron-positron pairs.  We refer to Sections 2 and 3 of Paper I for a more 
detailed description of these processes.  
Approximate values of the Coulomb pair creation cross-section at magnetic fields 
$B\geq B_{c}$ are given in Table 3 of Jones (2010b) and are the basis for the 
values of $W_{p}$ given here in Table 1.  These are not easily summarized as a 
simple expression analogous with equation (5). The cross-section at $B = 
10B_{c}$ is at least an order of magnitude smaller the the modified zero-field 
Bethe-Heitler cross-section but the effect on $W_{p}$ is not large because the 
photon track length in this region is limited by Compton scattering.  
Substitution into equation (4) gives values of $T_{max}$ that typically are 
reduced by a factor of approximately $0.9$ from those of equation (6).  But the 
complexity of the second-order processes at $B\sim B_{c}$ is such that we have 
not reconsidered shower development and, specifically, have not allowed for the 
production by cyclotron emission or Coulomb bremsstrahlung of photons with 
$k_{\perp}$ above the thresholds for magnetic conversion.  This becomes 
significant at $B = 10B_{c}$, as shown in Tables 2 and 4 of  Jones (2010b), and 
the high magnetic conversion transition rates reduce GDR-band photon track 
length in the shower.  The reduced $W_{p}$ values increase $T_{max}$.  Owing to 
this complexity at $B\sim B_{c}$ there are inevitable uncertainties in our 
estimates of $W_{p}$, but we believe that they do not seriously invalidate the 
estimates of the minimum polar-cap magnetic flux densities needed to support the 
case (iii) boundary condition and our conclusion that it can exist only in a 
very small subset of pulsars.  Our conclusions drawn from equation (3) are also 
independent of the spectrum of the reverse electron-photon flux because $W_{p}$ 
is almost completely independent of primary shower energy provided that is large 
compared with the GDR energy.

There appear to be no published estimates of the melting temperature of 
condensed matter that are specific to very high magnetic fields. Consequently, 
we are obliged to adopt the standard one-component Coulomb plasma expression 
(see, for example, Slattery, Doolen \& DeWitt 1980) which, with equation (1), 
gives,
\begin{eqnarray}
T_{m} = 1.0\times 10^{4}Z^{2}_{v}Z^{-0.23}B^{0.4}_{12}\hspace{5mm}{\rm K},
\end{eqnarray}
in terms of an effective valence charge $Z_{v}$. This latter parameter 
represents the fact that the deeply-bound Landau states certainly do not 
participate significantly in the melting transition, but its estimation at 
higher values of $Z$ is quite difficult. It is possible that the work of 
Potekhin, Chabrier \& Yakovlev (1997; see Fig. 1) could provide guidance 
although it is at zero field and was directed toward a different problem. On 
this basis, we assume for $Z = 26$ a value in the interval $Z_{v} = 10 - 15$.  
In a typical polar-cap field, $B_{12} = 10$, the melting temperature is as low 
as $T_{m} = 6\times 10^{5}$ K for $Z_{v} = Z = 6$ but exceeds $10^{6}$ K for $Z 
= 26$.  Thus the state of the polar cap may be a sequence of melting and 
solidifications.  The order of magnitude of the shear modulus is a further 
source of complexity. The standard (zero-field) expression for
a body-centred cubic lattice (Fuchs 1936) can be adapted, with equation (1), to 
give,
\begin{eqnarray}
\mu = 1.1\times 10^{16}Z^{2}_{v}Z^{-0.93}B^{1.6}_{12}\hspace{5mm}{\rm erg} 
\hspace{2mm}{\rm cm}^{-3}.
\end{eqnarray}
However, the polar-cap gravitational constant is $g\approx 2\times 10^{14}$ cm 
s$^{-2}$ so that any density inversion may well induce a form of Rayleigh-Taylor 
instability.

The final condensed-matter parameter that is important is the thermal 
conductivity parallel with ${\bf B}$, which is extremely high (see Potekhin 
1999).  Thermal energy is deposited at shower maxima a distance $z_{p}$ below 
the surface $z = 0$, which is defined as that separating condensed matter from 
the atmosphere,
 and is then dissipated as blackbody radiation from the polar cap.  The value of 
$z_{p}$ in the high-density condensed matter of the neutron-star surface depends 
on shower energy owing to the Landau-Pomeranchuk-Migdal (LPM) effect but for the 
order of magnitude of the characteristic time we can assume $-z_{p} \sim 10l_{r} 
\approx 10^{-3}$ cm 
using the radiation lengths $l_{r}$ given by equation (2) or in Table 1 of Paper 
I. The characteristic time for shower energy input to produce a 
surface-temperature fluctuation is then,
\begin{eqnarray}
t_{cond} = \frac{Cz^{2}_{p}}{2\lambda _{\parallel}} \approx 
10^{-9}\hspace{5mm}{\rm s},
\end{eqnarray}
in which typical values of the parameters are the specific heat $C = 1.0\times 
10^{12}$ erg cm$^{-3}$ K$^{-1}$ and the longitudinal coefficent of thermal 
conductivity
$\lambda _{\parallel} = 6\times 10^{14}$ erg cm$^{-1}$ s$^{-1}$ K$^{-1}$.
But for a surface temperature of $10^{6}$ K, the internal temperature gradient
needed to conduct the radiated energy flux
is extremely small, of the order of $10^{5}$ K cm$^{-1}$, within the condensed 
matter at  $z < 0$. In effect, heat is more easily conducted to greater depths 
than radiated from the surface.  Consequently, the Green function $G(z,z_{p},t)$ 
giving the internal temperature distribution must be almost independent of 
$z_{p}$ and very close to that for zero temperature gradient at the surface $z = 
0$.  Thus a shower heat input $H\delta(t)$ produces a temperature distribution 
within the condensed matter at $z \leq 0$,
\begin{eqnarray}
T(z,t) = HG(z,t) \approx \frac{H}{(\pi C\lambda _{\parallel}t)^{1/2}}\exp\left(
\frac{-Cz^{2}}{4\lambda _{\parallel}t}\right)
\end{eqnarray}
It is asymptotically $\propto t^{-1/2}$ so that the polar-cap temperature 
arising from a sequence of heating events, each producing a maximum temperature 
$T_{max}$, has fluctuations away from $\bar{T}$ whose magnitude is a function of 
the time-scale concerned.  Further discussion of this is deferred until our 
consideration of observed polar-cap blackbody temperature in Section 5.

\section[]{Short time-scale instability}

Under the assumption that the case (ii) boundary condition is satisfied, the 
neutron-star surface has an atmosphere in local thermodynamic equilibrium (LTE) 
with approximate scale height $l_{A} = (\tilde{Z} + 1)k_{B}T/Mg \sim 10^{-1}$ cm 
at temperature $T = 10^{6}$ K, where $\tilde{Z}$ and $M$ are the mean ion charge 
and mass of the partially ionized atom.
The expression for the chemical potential of an ideal Boltzmann gas gives the 
LTE atmospheric number density at $z = 0$,
\begin{eqnarray}
N_{A}(0) = \left(\frac{Mk_{B}T}{2\pi \hbar^{2}}\right)^{3/2}
{\rm e}^{-E_{c}/k_{B}T},        \nonumber    
\end{eqnarray}
for atmospheric temperatures such that $N_{A}(0) \ll N$. Its order of magnitude  
is $N_{A}(0) \approx 10^{32}\exp(-E_{c}/k_{B}T)$ cm$^{-3}$ for $M = 56m_{p}$.
We estimate the critical temperature $T_{c}$ which must not be exceeded if the 
case (iii) boundary condition is to be valid
by equating the flux of ions in such an LTE atmosphere that are incident on the 
neutron-star surface with the ion flux needed to give an open magnetosphere 
current density equal to $\rho_{0}c$. This gives $N_{A}(0) \approx 10^{14}$ 
cm$^{-3}$ and $k_{B}T_{c} = 0.025E_{c}$.  But the extent to which the atmosphere 
can be described as thin is very temperature-dependent. Thus for $T = 2T_{c}$ 
and $Z = 26$, $N_{A}(0)$ is of the order of $10^{23}$ cm$^{-3}$ and the whole 
atmosphere contains ions equivalent to about $10^{-1}l_{r}$.  At a density
$N_{A}(0) > 10^{23} - 10^{24}$ cm$^{-3}$, the Boltzmann gas estimate of $l_{A}$
assumed here is unreliable and it becomes necessary to allow for the interaction 
of protons with the ion atmosphere.

If instabilities in plasma acceleration with time-scales as short as$\sim 
10^{-4}$ s are considered, the temperature at $z = 0$ is constant apart from 
very small fluctuations, as shown by the Green function given at the end of the 
previous Section. Consequently, the temperature distribution and number density 
of ions in the atmosphere are also constant in time and the atmosphere is in 
local thermodynamic equilibrium in the interval $0 < z < z_{1}$, where $z_{1}$ 
is the top of the ion atmosphere, defined as the surface of last scattering. For 
time-scales of the order of $1 s$, the temperature and depth of the atmosphere 
can change appreciably but there is no doubt that it is always in local 
thermodynamic equilibrium.
The processes of proton formation by GDR decay
followed by diffusion to the surface were described in Paper I, which assumed a 
very thin atmosphere. Therefore, we need to consider here, in more detail, 
diffusion in the atmosphere $0 < z < z_{1}$. The number of protons is very many 
orders of magnitude smaller than that of ions so that the properties of the 
atmosphere, its scale height and equilibrium electric field given by the 
electrical neutrality condition, are determined solely by the latter.
With the assumption of a single ion component we can determine the equilibrium 
electric field within the LTE atmosphere in the presence of a gravitational 
acceleration $g$ and find that the proton potential energy is,
\begin{eqnarray}
\left(1 - \frac{A}{\tilde{Z} + 1}\right)m_{p}gz = \alpha m_{p}gz
\end{eqnarray}
for ions of charge $\tilde{Z}$ and mass number $A$, and that this
transports protons to $z > z_{1}$ from which region they are accelerated to 
relativistic energies in preference to the ions.  Proton diffusion at a rate 
greater than is needed for a current density $J^{p} = \rho _{0}c$ therefore cuts 
off ion acceleration and produces a thin electrically neutral proton atmosphere 
at $z > z_{1}$.  There appears to be no reason why this atmospheric structure 
should be disturbed by turbulent mixing.

The proton average potential energy at $z < 0$ is close to $m_{p}gz$ but the 
jump bias, $m_{p}ga_{s}/k_{B}T$ for ion separation $a_{s}$, is too small to be 
significant because $m_{p}gz_{p}/k_{B}T \ll 1$.  The atmospheric proton density 
at $z = 0$ is necessarily some orders of magnitude smaller than the density at 
$z < 0$ owing to the density discontinuity at the surface.
Thus diffusion to the surface at $z = 0$ is little changed by the presence of an 
atmosphere that is not necessarily very thin.
Movement of the protons through the ion atmosphere is effected by the chemical 
potential gradient which is a function of ion number density. At high densities, 
as in the solid at $z < 0$, this is determined principally by entropy but as the 
density reduces, the potential energy given by equation (12) becomes the more 
important factor.  In effect, the motion changes from a random walk to a drift 
velocity. The dividing ion density is given by the condition
$\lambda _{R} = N_{A}^{-1/3}$ and is
$\approx 4\times 10^{22}$ cm$^{-3}$, below which it is appropriate to define a 
drift velocity determined by the local value of the mean free path $\lambda 
_{R}$ derived from the total cross-section for Rutherford back-scattering,
\begin{eqnarray}
\bar{v} \approx -\alpha g\lambda _{R}\left(\frac{m_{p}}{k_{B}T}\right)^{1/2},
\end{eqnarray}
in which,
\begin{eqnarray}
\lambda _{R} = \frac{1}{\pi N_{A}}
\left(\frac{k_{B}T}{\tilde{Z}e^{2}}\right)^{2}.
\end{eqnarray}
The drift velocity is $\bar{v} \approx 2.6$ cm s$^{-1}$ at this density and 
$T = 10^{6} K$.
The consequence for a $T = 2T_{c}$ atmosphere is that the diffusion time to $z = 
z_{1}$ is increased, by some orders of magnitude, compared with the values 
assumed in Paper I which were for diffusion through the condensed matter only. 
Also, the change from random walk to drift velocity in the atmosphere has a 
significant effect on the distribution  of diffusion times because it removes 
the long tail in the distribution that is a feature of random walks.

Some effort has been made here to confirm that there is outward proton diffusion 
because the process is of crucial importance.
It was shown in Paper I that
plasma acceleration is unstable, consisting of alternate bursts of proton and 
ion acceleration, with time-scale determined by the time for proton diffusion 
from $z = z_{p}$ to $z = z_{1}$.  The polar-cap current density ${\bf J}$ is 
essentially time-independent for relativistic flow (as is the charge density) so 
that there is no fluctuation in polar-cap electric field other than that arising 
from the reverse flow of photo-electrons from accelerated ions (see Section 5).  
The instability is in the nature of the plasma accelerated and remains 
adequately described by the analysis given in Paper I which will not be repeated 
here.

The maximum acceleration potential difference, given a current density $J  = 
\rho _{0}c$, is almost entirely dependent on the Muslimov \& Tsygan (1992) 
general-relativistic correction.  There is a separate contribution arising from 
ion inertia (Michel 1974) but it is important only at very low altitudes. 
Neglecting this, the maximum potential difference is approximately
\begin{eqnarray}
|\Phi _{max}| \approx\frac{2\pi^{2}\kappa B_{d}R^{3}}{c^{2}P^{2}} \approx 
7\times 10^{12}
\frac{\kappa B_{d12}}{P^{2}}\hspace{5mm} {\rm volts}
\end{eqnarray}
and, unless restricted by pair creation at lower altitudes, is reached at an 
altitude $z$ of the order of the neutron-star radius $R$ which is roughly two 
orders larger than the polar-cap radius. We refer to Harding \& Muslimov (2001, 
2002) for a complete account.  We assume here that for this current density, 
spontaneous pair creation by curvature radiation is not possible.  Then the 
primary source of any positron component in ${\bf J}$ can only be the reverse 
flow of photo-electrons from accelerated partially-ionized atoms as discussed in 
Paper I. But there is an essential difference here in that protons in the very 
low density region at $z\sim z_{1}$ are almost completely ionized (here, we 
refer to Fig.1 of Potekhin et al 2006) so that both track length and energy flux 
of reverse-flow electrons from photo-ionized hydrogen atoms are negligibly 
small. Thus positron production in any interval of proton acceleration is 
negligible.  This is also true for ions of low atomic number $Z \sim 4-5$ which 
are completely ionized, but for higher $Z$, the reverse flux of inward 
accelerated electrons produces polar-cap ICS photons, as would outward 
accelerated electrons in case (i).   Pairs are produced by photons that are 
scattered to transverse momenta above the magnetic conversion thresholds.  The 
only difference is that the photons are inward moving.  But even if spontaneous 
CR pair formation is not possible,
positrons accelerated outward will produce CR pairs at higher altitudes, as in 
case (i), though superimposed on a flux of ions.

In case (i), the reverse flux of positrons must form protons which are not 
accelerated but form an atmosphere at $z > 0$ whose equilibrium is defined by 
various diffusion processes perpendicular to the magnetic flux.  Backward moving 
photons from the electron or positron showers are a further source of pair 
creation in each of cases (i) - (iii).  These arise from the decay of residual 
nuclei following proton or neutron emission in GDR formation, and from 
$(n,\gamma)$ reactions.  Those photons that are not absorbed in the more dense 
part of the atmosphere can produce pairs if their transverse momenta exceed the 
threshold.  The LPM effect, which in the high-density condensed matter at $z < 
0$ is significant for electrons of more than $10$ GeV or photons of more than 
$2$ GeV, is less important in the atmosphere within which there will be some GDR 
formation.  But we have been unable to make a satisfactory quantitative estimate 
of pair formation through this process.

\section[]{Medium time-scale instability}

Instability on time-scales some orders of magnitude longer than those of Section 
3 is also of interest and can appear as a fluctuation in the charge of nuclei 
reaching the polar-cap surface which we shall define here to be always at
$z = 0$ although it may move with respect to coordinates fixed at the centre of 
the star.
Ions of initial charge $Z_{a}$ move upward through the region of the shower 
maxima at $z_{p}$ and, with nuclear charge reduced to $Z_{b}$ by GDR formation 
and decay, enter the atmosphere at $z > 0$. We wish to find if there are 
conditions under which the local average nuclear charge $Z_{0}(z)$ fails to be a 
time-independent function of position with limits $Z_{a} \geq Z_{0}(z) \geq 
Z_{b}$.  The work of this Section replaces that of Section 4.1 in Paper I which 
contains an error.
  
The longer time-scales here enable us to assume the proton and ion current 
densities $J^{p}$ and $J^{z}$ are the time averages of those described in the 
previous Section.  There is, of course, a distribution of discrete nuclear 
charges in the shower region, but we shall work in terms of the average 
$Z(z,t)$, the corresponding number density $N(z,t)$ given by equation (1), and 
the velocity $v(z,t)$ with which nuclei approach the polar-cap surface at $z = 
0$.  Proton formation by GDR decay occurs predominantly in the very late stages 
of shower development, as explained in Section 1, and so we shall assume that it 
is confined within limits $z_{a}$ and $z_{b}$ and is defined by the normalized 
function $g_{p}(z)$,
\begin{eqnarray}
\int^{z_{b}}_{z_{a}}g_{p}(z)dz = 1.
\end{eqnarray}
The physical basis for our study of the system is that the total numbers of 
nuclei and of protons (bound or unbound) are conserved. Because both neutrons 
and protons are produced in GDR decay, we can make the approximation of 
neglecting the effect of $\beta$-transitions.  Therefore, 
the continuity equations are,
\begin{eqnarray}
\frac{\partial N}{\partial t} + \frac{\partial(Nv)}{\partial z} = 0,
\end{eqnarray}
and with neglect of the relatively short proton diffusion time so that within 
medium time-scales all protons produced in the shower are assumed to be promptly 
accelerated from the atmosphere at $z \approx z_{1}$,
\begin{eqnarray} 
\frac{\partial (NZ)}{\partial t} + \frac{\partial (NZv)}{\partial z}
 = -g_{p}(z)J^{p}(t).
\end{eqnarray}
Because the reverse electron flux from photo-electric ionization is the source 
of the showers, we shall find it convenient here, as in Paper I, to introduce 
the parameter $K$ which is a function of the atmospheric nuclear charge and is 
the number of protons produced per unit positive nuclear charge accelerated.  
With the representation of equation (1) in the form
$N = CZ^{\gamma}$, equations (17) and (18) can be combined to give,
\begin{eqnarray}
C\int^{z_{b}}_{z_{a}}dz\left(Z^{\gamma}\frac{\partial Z}{\partial t} +
vZ^{\gamma}\frac{\partial Z}{\partial z}\right) = -J^{p}(t).
\end{eqnarray}
It has the obvious time-independent solution,
\begin{eqnarray}
Z_{a} - Z_{b} = KZ_{b},
\end{eqnarray}
and the time-independent velocity of nuclei is such that
$Z^{\gamma}_{0}(z)v_{0}(z)$ is independent of $z$.

A natural fluctuation away from $Z_{0}$ would be of the form $Z(z,t) = Z_{0} + 
\delta Z(z,t)$ with,
\begin{eqnarray}
\delta Z = \eta\left(Z_{a} - Z_{0}(z)\right){\rm e}^{{\rm i}\omega t},
\end{eqnarray}
giving similar fluctuations in $J^{p}$, $N$ and $v$, in which $\eta$ is 
infinitesimal and independent of $z$ and $t$.
From equation (17) we then have,
\begin{eqnarray}
\delta\left(Z^{\gamma}v\right) = -{\rm i}\omega\eta\chi(z){\rm e}^{{\rm i}\omega 
t},
\end{eqnarray}
where,
\begin{eqnarray}
\chi(z) = \gamma\int^{z}_{z_{a}}dz^{\prime}\left(Z_{a} - Z_{0}\right)Z^{\gamma 
-1}_{0}.
\end{eqnarray}
Substitution into equation (19) with the retention of terms of first order in 
$\eta$ gives,
\begin{eqnarray}
{\rm i}\omega\eta {\rm e}^{{\rm i}\omega 
t}\int^{z_{b}}_{z_{a}}dz\left(Z^{\gamma}_{0}\left(Z_{a} - Z_{0}\right) - 
\chi(z)\frac{\partial Z_{0}}{\partial z}\right) -  \nonumber  \\ 
Z^{\gamma}_{0}v_{0}\left(Z_{b} - Z_{a}\right)
\eta {\rm e}^{{\rm i}wt} = - \frac{1}{C}\delta J^{p}(t).
\end{eqnarray}
The current densities, averaged over short time-scales, are $J^{z} = NZv(0,t)$ 
and, from the definition of $K$, $J^{p} = KJ^{z}$.
The fluctuation away from the time-independent solution is $\delta J = \delta 
J^{p} + \delta J^{z} = 0$,
which we assume to be
maintained by the boundary conditions on $\Phi$. To express
$\delta J^{p}$ in terms of $\delta Z$ we represent the $Z$-dependence  of $K$ in 
the vicinity of $Z_{b}$ as a power law $K = K_{0}Z^{\nu}(0,t)$.
We find, after eliminating the velocity fluctuation $\delta v(0,t)$, that
\begin{eqnarray}
\frac{1}{C}\delta J^{p} = \nu Z^{\gamma}_{0}v_{0}\frac{K}{K + 1}\delta Z(0,t).
\end{eqnarray}
The relationship with equation (24) is established by noting that,
\begin{eqnarray}
\delta Z(0,t) = \delta Z(z_{b},t-\tau)  
	  = \eta\left(Z_{a} - Z_{b}\right){\rm e}^{{\rm i}\omega(t-\tau)},
\end{eqnarray}
where $\tau$ is the time interval of nuclear movement from $z = z_{b}$ to $z = 
0$.  From equations (24) - (26), the equation whose root $\omega$ we require can 
be expressed as,
\begin{eqnarray}
\frac{{\rm 
i}\omega}{Z^{\gamma}_{b}v_{0}(z_{b})}\int^{z_{b}}_{z_{a}}dzZ^{\gamma-1}_{0}
\left((1 + \gamma)Z_{0} - \gamma Z_{b}\right)\left(Z_{a} - Z_{0}\right)
\nonumber   \\
 = - (Z_{a} - Z_{b}) - \frac{\nu K}{K + 1}(Z_{a} - Z_{b})
{\rm e}^{-{\rm i}\omega \tau},
\end{eqnarray}
in which the function $\chi(z)$ has been removed by partial integration.

The only assumptions we need make about the depth distribution of proton 
formation in the late stage of shower development are that it is small outside 
the interval $z_{a} < z < z_{b}$ and that $z_{b} - z_{a}$ is smaller than 
$|z_{b}|$ though not necessarily so by as much as an order of magnitude.  A 
suitable function would be,
\begin{eqnarray} 
g_{p}(z)) = \frac{2}{z_{b} - z_{a}}\sin^{2}\left(\frac{\pi (z - z_{a})}
{z_{b} - z_{a}}\right),
\end{eqnarray}
in terms of which we could express $Z_{0}$ as an explicit function of $z$,
which would be needed to obtain numerical values for the root $\omega = \omega 
_{1} + {\rm i}\omega _{2}$ of equation (27).  
However, we are primarily concerned here not with growth rates but with the 
boundary between stability and instability and so shall not proceed with 
numerical solutions for $\omega$.
Fortunately, it is possible to obtain a sufficient condition for the existence 
of instability independently of the form of $g_{p}$.   Provided $-1 < \gamma < 
0$, which is clearly satisfied, we can see directly from the values of the 
integrand in equation (27) at the limits and from its lack of an extremum that,
\begin{eqnarray}
\int^{z_{b}}_{z_{a}}dz\left((1 + \gamma)Z_{0} - \gamma Z_{b}\right)
\left(\frac{Z_{a} - Z_{0}}{Z_{0}}\right)
\left(\frac{Z^{\gamma}_{0}}{Z^{\gamma}_{b}}\right)  \nonumber   \\
< (z_{b} - z_{a})(Z_{a} - Z_{b}),
\end{eqnarray}
and therefore that equation (27) can be replaced by the inequality,
\begin{eqnarray}
{\bf i}\omega\tau\left(\frac{z_{a} - z_{b}}{|z_{b|}}\right) > 
-1 - \frac{\nu K}{K + 1}
{\rm e}^{-{\rm i}\omega\tau}.
\end{eqnarray}
Thus$\omega _{1,2}$ satisfy the inequalities,
\begin{eqnarray}
\omega _{1}\tau\left(\frac{z_{b} - z_{a}}{|z_{b}|}\right) & > &
\frac{\nu K}{K + 1}{\rm e}^{\omega _{2}\tau}\sin\omega _{1}\tau
\nonumber   \\
\omega _{2}\tau \left(\frac{z_{b} - z_{a}}{|z_{b}|}\right) & < &
1 + \frac{\nu K}{K + 1}{\rm e}^{\omega _{2}\tau}\cos\omega _{1}\tau.
\end{eqnarray}
From the first of these, we can see that the real part of $\omega$ must satisfy 
$\omega _{1}\tau > \pi/2$, provided the proton formation region in the shower is 
sufficiently compact that,
\begin{eqnarray}
\frac{\nu K}{K + 1}{\rm e}^{\omega _{2}\tau} > \frac{\pi}{2}\left(\frac
{z_{b} - z_{a}}{|z_{b}|}\right).
\end{eqnarray}
With greater compactness, $\omega _{1}\tau \rightarrow \pi$, as would be 
expected.  The second inequality gives the condition for $\omega _{2} < 0$, that 
is, fluctuation growth. At the threshold where $|\omega _{2}\tau| \ll 1$, and 
for $z_{b} - z_{a} \ll |z_{b}|$, it is simply $\nu K > K + 1$.

The $Z$-dependence of $K$ was represented in Paper I as an approximate power
law, $K = K_{0}Z^{\nu}$ with $\nu = 0.85$ for $Z \geq 10$ but with the 
reservation that this would certainly become invalid for an atmosphere of ions 
with $Z \sim 5$ which would be almost completely ionized.  Therefore, we should 
anticipate quite large values of $\nu$ at small $Z$, sufficient to give the 
unstable behaviour found here.  Stability clearly depends on the energy flux 
from photo-ionization.  Low fluxes and moderate rates of proton formation in the 
GDR region of the showers, such that the nuclear charge inferred from equation 
(20) is $Z_{b} \approx 10$, allow a stable time-independent progression of 
nuclear charge as a function of depth below the polar-cap surface as given by 
$Z_{0}(z)$.  But larger values of $K$ and
smaller $Z_{b}$ lead to instability. (The value of $Z_{a}$ is unimportant, 
provided it is not small, and for order of magnitude estimations in this paper 
we have assumed $Z_{a} = 26$ unless otherwise stated.)

Unfortunately, our analysis of the instability of nuclear movement to the 
polar-cap surface is limited to small fluctuations and does not extend to large 
amplitudes. But it is not difficult to see the form it would take.  An 
atmosphere of high-$Z$ ions at time $t$ produces a high reverse-electron energy 
flux which creates a layer of low-$Z_{b}$ ions in the shower maximum region 
$z_{a} < z < z_{b}$. These flow toward the surface at $z = 0$ and form a low-$Z$ 
atmosphere at time $t + \tau$ which produces a low reverse-electron energy flux 
and correspondingly large values of $Z_{b}$.  The ions accelerated alternate 
between high and low-$Z$ values.  The basic unit of time is given by the time 
$t_{rl}$ for the emission at the Goldreich-Julian current density of one 
radiation length of ions,
\begin{eqnarray}
t_{rl} = 2.1\times 10^{5}Z^{-1}B^{-1}_{12}(- P\sec \psi) \nonumber  \\
\left(
\ln \left(12Z^{1/2}B^{-1/2}_{12}\right)\right)^{-1}\hspace{5mm}{s}.
\end{eqnarray}
The high-$Z$ intervals are subject to short time-scale instability as described 
in Section 3, but low-$Z$ intervals have little or no reverse-electron flux and 
therefore no possibility of significant positron acceleration and 
electron-positron pair production.

The instability outlined here is of quite simple form, but there are  
complicating factors that have been mentioned earlier in the Section 2 
consideration of polar-cap parameters.  Evaluations of the melting temperature 
discussed following equation (8) show that there is every possibility that the 
condensed matter state below the atmosphere may be liquid, or a solid close to 
melting with a high self-diffusion rate. This would have no effect on short 
time-scale instability but could complicate the behaviour of the system over 
medium time-scales.  The melting temperature is $Z$-dependent so that a density 
inversion is possible, the upper layer of higher-$Z$ being either liquid or 
solid.  In the liquid case, we must anticipate Rayleigh-Taylor instability, 
which may also exist in the solid case because its shear modulus (see equation 
9) may not be adequate to maintain mechanical stability.  All these processes 
are occurring at depths $|z_{p}| \sim 10^{-3}$ cm but over a polar cap whose 
radius is of the order of $10^{4}$ cm. Consequently, a further complication is 
that different polar-cap areas are unlikely to be in phase with each other.  
These problems are considered further, though necessarily in a qualitative way, 
in Section 5.  

\section[]{Nulls subpulse drift and polar-cap coherence}

The time-dependent phenomena considered in Sections 3 and 4 are local and 
one-dimensional because both shower depth $z_{p}$ and atmospheric scale height 
are very many orders of magnitude smaller than the polar-cap radius. This 
introduces the question of whether or not there is coherence over the
whole polar-cap area. Both instabilities are dependent on the parameter $K$
which is a function of the surface nuclear charge $Z(0,t)$ and to a lesser 
extent of surface temperature and acceleration field. For this reason alone, we 
see no case for assuming complete coherence and anticipate that the polar cap we 
describe has zones of proton and ion emission which cannot be stationary, the 
total areas of each being determined, approximately, by the average value of 
$K$. For neutron stars that are unable to support spontaneous pair production by 
curvature radiation, the proton zones have no reverse electron flux, do not 
support pair production, and hence merely produce an accelerated one-component 
plasma as described in Section 3.  But the ion zones will support ICS pair 
production and so appear as moving sources within the polar cap. In this 
Section, we shall attempt to compare possible forms this motion might take with 
some of the observed phenomena in radio pulsars.

The distinction between the average pulse profile and individual sub-pulses 
within it was noted almost immediately following the discovery of pulsars
(see, for example, Smith 1977).  The amplitude and form of successive sub-pulses 
can vary in times of the order of the rotation period and in some pulsars, 
observed with higher resolution, sub-pulses have micro-structure of $10^{-4}- 
10^{-3}$ s time-scale. There are also more organized phenomena, and for recent 
extensive surveys of these we refer to Wang et al (2007) in the case of pulse 
nulls and to Weltevrede, Edwards \& Stappers (2006) for sub-pulse drift. There 
is a move toward a consensus (Lyne et al 2010) that quasi-periodic switching 
between magnetospheric states with different spin-down rates is the basis of 
mode-changing. These authors even suggest that it is the source of a large 
component of pulsar timing noise. 
However, it is also true that the subpulse characteristics observed in a
small number of pulsars are quite singular, but here discussion is confined to 
the general features of subpulses.

Given the properties of the medium time-scale instability described in Section 
4, it is natural to associate with nulls those intervals in which high-$Z$ ion 
zones either do not exist or are confined to areas of the polar cap from which 
radiation produced by the plasma is not visible to the observer.  For 
sufficiently low values of $Z_{b}$, ions are accelerated from the surface 
completely ionized so that there is no reverse-electron flux and hence no pair 
creation and radio emission.  The current density is little changed in the open 
sector of the magnetosphere but the absence of pair creation means that the 
particle content near the light cylinder is quite different as are the 
components of the momentum density or stress tensor on any spherical surface in 
this region centred on the star.  It is therefore not surprising that the 
spin-down torque is reduced during the interval of a null, as has been observed 
in PSR B1931+24 (Kramer et al 2006).  Neutron stars with small $K$ such that 
$Z_{b} \approx 10$ are likely to have $\nu < 1$ and so will have a steady-state 
progressively reducing nuclear charge $Z_{0}(z)$, no medium time-scale 
instablity, and no long-term nulls.  But the questions about melting and 
mechanical stability of the polar cap which were mentioned at the end of Section 
4 remain and the whole system is so complex and difficult to describe in 
physical terms that we are unable to give useful quantitative predictions. 
However, incomplete nulls, having a low but detectable level of emission should 
be observed.  It is also unsurprising that nulls are observed to be not 
completely random (Redman \& Rankin 2009).

To some extent and unfortunately, the same remarks have to be made about the 
effects of short time-scale instability. For temperatures $T > 2T_{c}$, 
the proton diffusion time is much longer than in the case of the very thin 
atmosphere assumed in Paper I but remains difficult to calculate with complete 
confidence. In order to describe the polar cap it is necessary to introduce 
coordinates ${\bf u}(z)$ on a surface perpendicular to ${\bf B}$.
As in Paper I, the proton current density at any point ${\bf u}(0)$ can be 
related to the ion current density through the definition of $K$,
\begin{eqnarray}
J^{p}({\bf u},t) + \tilde{J}^{p}({\bf u},t) = 
K\int^{t}_{-\infty}dt^{\prime}f_{p}(t - t^{\prime})J^{z}({\bf u},t^{\prime}),
\end{eqnarray}
in which $f_{p}$ is the normalized proton diffusion-time distribution.  Without 
the $\tilde{J}^{p}$ term, whose significance is described below, this would be a 
homogeneous Volterra equation of the second kind having no non-zero 
square-integrable solution. (The time-dependence of $K$ arising from the 
temperature dependence of the LTE ionic charge $\tilde{Z}$ is neglected here.)
The approximate expression for $f_{p}$ given in Paper I (equation 23) assumed a 
random-walk diffusion through the condensed matter at $z < 0$ and so is not 
valid for an atmosphere with properties given by equations (13) and (14).
For $T > 2T_{c}$, most of the diffusion time is in the drift-velocity phase, in 
which case the time distribution would be more suitably approximated by a 
normalized gaussian function centred at $t - \tau _{p}$ or, in the limit, by
$f_{p}(t - t^{\prime}) = \delta(t - t^{\prime} - \tau _{p})$, where
$\tau _{p}$ is derived from equation (13) of Section 3.
The quantity $\tilde{J}^{p} = 0$ within intervals for which $J^{p} < 1$ and at 
other times, when $J^{p} = J$, represents the storage of excess protons in the 
atmosphere at $z > z_{1}$.  The total current density $J$ is fixed by the 
boundary conditions and at $z = 0$ differs little from the Goldreich-Julian 
value.  Thus $J^{p} = J$ and $J^{z} = 0$ until the instant at which the 
atmosphere is exhausted and $J^{p}$ falls to some residual value $J^{p} < J$.  
Then ion flow recommences almost immediately (in a time much shorter than the 
growth time for spontaneous curvature radiation pairs) and continues until 
proton diffusion grows sufficiently to re-form the proton atmosphere.  The 
durations of the time intervals for ion and proton emission are those for 
electron-positron pair creation or otherwise, and labelled $\tau _{ee}$ and 
$\tau _{gap}$ are given by,
\begin{eqnarray}
  J  & =  & K\int^{\tau _{ee}}_{0}dt^{\prime}f_{p}(\tau _{ee} - 
t^{\prime})J^{z}(t^{\prime})     \nonumber   \\
\tau _{gap} & = & \frac{K}{J}\int^{\tau _{ee}}_{0}dt^{\prime}J^{z}(t^{\prime})
\end{eqnarray}
Therefore $\tau _{ee}$ and $\tau _{gap}$ both depend on the smallness of $f_{p}$ 
for small values of $t - t^{\prime}$.  In the $\delta$-function limit for 
$f_{p}$, we have $\tau _{gap} = K\tau _{ee}$.  Estimates of the parameter
$K = K_{0}Z^{\nu}$ were given in Paper I, but are repeated here,
\begin{eqnarray}
K_{0} & = & 2.8\langle Z_{sm}\rangle^{-0.76}B^{0.62}_{12}T^{-1.0}_{6},
\hspace{1cm}B_{12} > 1,    \nonumber  \\
K_{0} & = & 1.6\langle Z_{sm}\rangle^{-0.76}T^{-1.0}_{6}, \hspace{1cm}
B_{12} < 1, 
\end{eqnarray}
with $\nu = 0.85$ for $10 < Z < 26$.
In these expressions, $T$ is not the local surface temperature but is an average 
for radiation emitted over the whole polar-cap, and $B$ is the actual polar-cap 
magnetic flux density.  For $B_{12} \gg 1$, the values of $K_{0}$ given need to 
be modified, though not greatly, to allow for the high-$B$ values of $W_{p}$ 
contained in Table 1.

Guided by the estimates contained in equations (6) and (7), we anticipate that 
except for a very small number of neutron stars, the surface temperature is at 
all times $T > T_{c}$ so that the case (ii) boundary condition is always 
maintained and ion emission is never temperature-limited. It is limited instead 
by the fact that the proton atmosphere forms above the ions and the protons are 
preferentially accelerated, as described in Section 3. The only effect of the 
temperature variations described by equation (11) that occur as a result of 
alternating proton and ion emission is to change the density of the LTE ion 
atmosphere. It is also worth observing that the local nature of equation (34) is 
not disturbed by the presence of ${\bf E}\times{\bf B}$ drift above the polar 
cap.  Although this slightly displaces a reverse electron shower relative to the 
point from which the ion was accelerated, it has no effect on ion emission, 
which is not temperature-dependent.

In view of the time-variation described by equations (34) and (35), is it 
possible to imagine organized rather than chaotic motion of ion zones on the 
polar cap?  An example of chaotic motion would be the existence of very many 
small zones without obvious organization.
Two simple organized cases would be motion along a slot of constant rotational 
latitude and circular motion at constant $u(0)$.  With the Deshpande \& Rankin 
(1999) analysis of PSR B0943+10 in mind, we consider circular motion.  Equations 
(34) and (35) are local in ${\bf u}$ and contain no information that can 
determine an ion zone movement velocity
$\dot{\bf u}$.  The quantities that are essentially constant are $K$ and, apart 
from the effect of varying LTE atmospheric temperature, $\tau _{p}$.  Thus the 
circulation time for $n$ ion zones, is $\hat{P}_{3} = nP_{3}$, where $P_{3}$ is 
the band separation in the usual notation 
by which subpulse drift is described, and here is given by
\begin{eqnarray}
P_{3}   =  \tau _{gap} + \tau _{ee} = (K + 1)\tau _{p} 
\end{eqnarray}	 
in the $\delta$-function limit of the diffusion function $f_{p}$.  In this 
system, the velocity $\dot{\bf u}$ is determined by $n$ for fixed values of $K$ 
and $\tau _{p}$.
For the same reason, drift direction is also unspecified and there is nothing to 
preclude a change in $n$ or a reversal following some disturbance to the 
polar-cap surface such as might be a consequence of the kind of mechanical 
instability briefly described in Section 4. The drift time is determined 
principally by diffusion through the more dense layers of the atmosphere.  From 
equations (13) and (14) it is,
\begin{eqnarray}
\tau _{p} = \int^{z_{1}}_{0}\frac{dz}{\bar{v}(z)} = \left(\frac{l_{A}}{\alpha 
g\lambda _{R}(0)}\right)\left(\frac{k_{B}T}{m_{p}}\right)^{1/2}.
\end{eqnarray}
In Section 3 we observed that atmospheric density at $z = 0$ is an exponential 
function of $T > T_{c}$ so that its value is essentially unpredictable.  
Therefore, it is likely that the density discontinuity at $z = 0$ is
small and that the reverse-electron showers may be contained entirely within the 
atmosphere.  In this case, the lower integration limit in equation (38) must be 
replaced by $z_{sm} > 0$.  The diffusion time is then not directly a function of 
$B$, but is dependent on surface gravity and on $\tilde{Z}$.  But its 
distribution for all pulsars should be compact.  The distribution in the values 
of $K$, given by equation (36), is probably the more important source of the 
observed spread in the values of $P_{3}$. 
Detailed calculation of $\tau _{p}$ has not been attempted here. In particular, 
our estimates of $\lambda _{R}$ and $l_{A}$ are subject to some uncertainty.

The fact that $P_{3}$ is constant for a particular pulsar whereas the 
circulation time $\hat{P}_{3}$ is dependent on $n$ could in principle allow 
comparison with ${\bf E}\times {\bf B}$ drift-velocity polar-cap models in which 
$\hat{P}_{3}$ is constant.  It is also worth considering the effect of 
mechanical instability in allowing leakage of protons (or low-$Z$ ions) to 
limited areas of the surface as described at the end of Section 4.  The excess 
protons form a localized atmosphere of greater depth than in surrounding areas, 
so that ion zones (with consequent pair creation) do not form there until it is 
exhausted.  The observer of radio emission probably sees plasma from no more 
than a strip of polar-cap area at roughly constant rotational latitude so it is 
possible that intervals of null emission would be
be observed at fixed longitude in a sequence of subpulse bands.  

In a survey of $187$ pulsars, selected only by signal-to-noise ratio,
Weltevrede et al (2006) found that reversals do occur and that roughly   equal 
numbers of pulsars have subpulse drift to smaller or greater longitudes. They 
also comment that subpulse drift is so common a phenomenon that it cannot depend 
on pulsar parameters having extraordinary values.
The band separation $P_{3}$ is independent of radio frequency, and is not 
correlated with period $P$, age $P/2\dot{P}$, or with the inferred dipole field 
$B_{d}$.  Measured values of $P_{3}$ given for 77 pulsars in Table 2 of 
Weltevrede et al are mostly contained within a single order of magnitude, $1 < 
P_{3} < 10$ s although the distribution has a small tail extending to $\sim 20$ 
s.  This is quite compact (for a neutron-star parameter) and is not inconsistent 
with the interpretation of subpulse drift given here and with equation (37).  
The range of $P_{3}$, with estimates of $K$ found from equations (36), indicate 
$\tau _{p} \sim 10^{-1} - 10^{0}$ s and $N_{A}(0) = 10^{23} - 10^{24}$ 
cm$^{-3}$, values that are by no means unreasonable.

It is necessary to compare our polar-cap model with two other sets of pulsar 
observations.  Given that protons are the major fraction of particles 
accelerated and produce no electron-positron pairs, it is worthwhile considering 
the radio luminosity $L_{\nu}$. The order of magnitude of this can be expressed 
as,
\begin{eqnarray}
L_{\nu}\Delta\Omega\Delta\nu \approx \frac{2\pi^{2}R^{3}B_{d}}{ceP^{2}}\epsilon
    \approx  \left(\frac{1.4\times 10^{30}B_{d12}}{P^{2}}\right)\epsilon
\end{eqnarray}
in terms of the neutron-star radius $R$ and the inferred dipole field $B_{d}$.
From this we can estimate the energy $\epsilon$ radiated into solid angle 
$\Delta\Omega$ within bandwidth $\Delta\nu$ per unit charge (baryonic or 
leptonic) accelerated at the polar cap.  Using the 400 MHz
luminosities listed in the ATNF catalogue (Manchester et al 2005) and a 
bandwidth $\Delta\nu = 400$ MHz, we find by evaluating equation (37) for a small 
sample (B0826-34, B0834+06, B0943+10, B0950+08, B1055-52, B1133+16, B1929+10) 
that typical values are in the interval
$\epsilon = (40 - 400)\delta\Omega$ MeV.  Even though the emission solid angle 
is likely to be as small as $\Delta\Omega \sim 10^{-2} - 10^{-1}$ sterad, there 
must be some concern here because the number of ICS pairs produced per primary 
accelerated positron (Harding \& Muslimov 2002) is not large so that either the 
conversion of electron-positron energy to coherent radio emission is efficient 
or other plasma components are involved, as in the paper by Cheng \& Ruderman 
(1980).

A second set of observations that are relevant are those of polar-cap blackbody 
X-ray luminosities.  The blackbody X-ray emission expected from a polar-cap of 
ion and proton zones can be found from equation (11). The reverse-electron flux 
from photo-ionization heats an ion zone at a rate $H_{0}$ within a time interval 
$0 < t < \tau _{p}$. An estimate of this can be obtained directly from the mean 
electron energy per unit nuclear charge accelerated, given by
equation (29) of paper I, and is,
\begin{eqnarray}  
H_{0} = 6.0\times 10^{18}Z^{0.85}(0)B^{1.5}_{12}T^{-1}_{6}P^{-1},
\end{eqnarray} 
in units of erg cm$^{-2}$ s$^{-1}$. In this expression, $T$ is not the local 
surface temperature but is an average for radiation emitted over the whole polar 
cap. The surface temperature derived from equation (11) with neglect of 
radiative loss is
\begin{eqnarray}
T(t) = \frac{2H_{0}}{(\pi C\lambda _{\parallel})^{1/2}}t^{1/2}, \hspace{5mm}0 < 
t < \tau _{p},
\end{eqnarray}
and increases very rapidly, so that the limiting temperature,
\begin{eqnarray}
T_{06} \approx 0.6Z^{0.17}(0)B^{0.3}_{12}P^{-0.2},
\end{eqnarray}
derived from $H_{0}$ is reached at $t \ll \tau _{p}$.  The cooling at $t > \tau 
_{p}$ is also rapid so that the major part of the X-ray luminosity is that of a 
black body at $T_{0}$ and of area approximately equal to the canonical 
dipole-field area $2\pi^{2}R^{3}/cP$ divided by $K + 1$.

There have been many attempts to measure the polar-cap blackbody temperature and 
source area of a subset of radio pulsars.  We refer to Zavlin \& Pavlov (2004) 
for B0950+08 ; De Luca et al (2005) for B0656+14 and B1055-52; 
Tepedelenlio\u{g}lu \& \"{O}gelman (2005) for B0628-28; Zhang, Sanwal \& Pavlov 
(2005) for B0943+10; Kargaltsev, Pavlov \& Garmire (2006) for B1133+16; Gil et 
al (2008) for B0834+06; Misanovic, Pavlov \& Garmire (2008) for B1929+10.
Five pulsars (B0628-28, B0834+06, B0943+10, B1133+16, B1929+10) have source 
areas one or two orders of magnitude smaller than the canonical area, but
unfortunately, a systematic comparison with equations (40)-(42) is not possible 
because in most instances the authors are able to say only that the observed 
X-ray spectrum is consistent with the stated temperature and source area.  The 
quoted source temperatures are $\sim 3\times 10^{6}$ K and are larger than those 
predicted by equation (42) unless it is assumed that $B \gg B_{d}$.   They are 
also uncomfortably large in the context of ${\bf E}\times {\bf B}$ 
drift-velocity polar-cap models such as that developed by Gil, Melikidze \& 
Geppert (2003) although it must be conceded that this model allows an 
interesting test of its validity (see Gil, Melikidze \& Zhang 2007).
However, the case (iii) surface electric-field boundary condition
on which these models rely could be maintained only for ion cohesive energies of 
$\sim 10$ keV which in turn would imply actual fields two orders of magnitude or 
more larger than the ATNF catalogue dipole field. 

\section{Conclusions}

This paper is a continuation of a previous study (Paper I) of isolated neutron 
stars with positive corotational charge density and surface electric-field 
boundary condition ${\bf E}\cdot{\bf B} = 0$ at the polar cap.  The reverse flux 
of electrons arising from photo-ionization of accelerated ions is incident on 
the neutron-star surface and produces protons through formation and decay of the 
giant dipole resonance in the later stages of electromagnetic shower 
development.  Protons are the major component of the accelerated plasma  but we 
find that a time-independent composition of ions, protons and positrons is 
usually unstable. The consequences of these phenomena should be observable in 
pulsars unless, of course, rendered nugatory by some factor not properly taken 
into account here.  But we emphasize that there are almost certainly other 
sources of instability  contributing to the complex behaviour observed in radio 
pulsars that may be present in each of cases (i) - (iii).

There are two instabilities which result in transitions between states of 
different plasma composition above localized areas on the polar cap and it is 
suggested here that these are the basis for the commonly observed phenomena of 
nulls and of subpulse drift.  The basic units of time are $\tau _{p} \sim 
10^{-1}$ s for the short time-scale instability and for the medium time-scale, 
$t_{rl} \sim 10^{2} - 10^{3}$ s. These instabilities are not primarily 
electromagnetic in origin and will not be present in neutron stars with negative 
corotational charge density and electron acceleration.  Micropulses of $\sim 
10^{-4} - 10^{-3}$ s duration are also present in some pulsar emissions but are 
unlikely to be associated with the instabilities we consider here.
There appears to be no reason why both signs of corotational charge density 
should not be present in the isolated neutron-star population and it would be 
interesting to see if there is any observational evidence for an associated 
division of the ATNF catalogue pulsars. 

The transitions are principally in plasma composition and the current density at 
the neutron star surface remains close to $J=\rho _{0}c$.  The electric 
potential $\Phi(z,{\bf u}(z))$ in the open magnetosphere above the polar cap is 
also almost unchanged.  These transitions are, in principle, observable because 
there are no processes by which electron-positron pair creation can occur in a 
plasma containing only protons, whereas it does occur in the ion plasma.   Thus 
there will be no coherent radio emission in the proton phase if we assume, as is 
usual, that a pair plasma is a necessary constituent for it.

Both the medium time-scale instability and the mechanical instability of the 
neutron-star surface mentioned in Section 4 provide a basis for pulse nulls, but 
unfortunately, not one that is suitable for quantitative prediction.  The short 
time-scale instability is more interesting.  We show that at any instant, the 
polar cap is divided into moving ion and proton zones and consider organized 
motion of these, specifically the circular movement of compact ion zones around 
the magnetic pole.  We make the hypothesis that such motions can occur 
(following recent work by Deshpande \& Rankin 1999) but
have not attempted to show that they have long-term stability, to the extent 
indicated by observation, against decay to a chaotic state.  This is of some 
interest because it shows that features usually considered to require the 
surface-field boundary condition ${\bf E}\cdot{\bf B} \neq 0$ are possible under 
the ${\bf E}\cdot{\bf B} = 0$
condition.  Actual polar magnetic flux densities exceeding $10^{14}$ G are 
necessary for the former condition and it must be doubtful that such fields, two 
orders of magnitude greater than the inferred (catalogue) dipole field, are 
likely to exist in the very large number of pulsars observed by Weltevrede et al 
(2006) to show subpulse drift.  Our view of subpulse drift is also interesting 
in that the motion is not an ${\bf E}\times {\bf B}$ drift velocity as in the 
model of Ruderman \& Sutherland.  The fixed parameters for a given neutron star 
are the band separation $P_{3}$, and $K$ which is the number of protons produced 
per unit ion charge accelerated.  This latter parameter approximately determines 
the ratio of the total areas of proton and ion zones on the polar cap at any 
instant.  Thus the circulation time $\hat{P}_{3} = nP_{3}$ depends on $n$, the 
number of ion zones, and is not necessarily constant as it would be in the ${\bf 
E}\times{\bf B}$ drift velocity model. The band separation $P_{3}$ is dependent 
on $K$ and on the diffusion time $\tau _{p}$ and so is independent of rotation 
period $P$ and almost independent of $B$.

We have assumed here, following the pair formation calculations of Hibschman \& 
Arons (2001) and Harding \& Muslimov (2002), that spontaneous pair creation by 
curvature radiation is not possible in most isolated neutron stars.  Those with 
a sufficiently large value of $|\Phi _{max}| \propto B_{d}/P^{2}$ to support 
spontaneous CR pair creation, predominantly young high-field pulsars, are not 
expected to show the instabilities we have considered here. These authors 
actually considered case (i) in which the basis of plasma formation is an 
electron current density close to $J = \rho _{0}c$ at the surface accelerated to 
energies sufficient for ICS or spontaneous CR pair formation.  In case (ii), the 
protons and ions form the basic current component with $|\Phi _{max}|$ broadly 
the same as in case (i) apart from the ion inertia term which is very
small for electrons.  Thus the condition for the spontaneous growth of CR pair 
formation is very similar in cases (i) and (ii).  Although the reverse electron 
flux arising from CR pair formation in case (ii) is probably quite a small 
fraction of $\rho _{0}c$, it is likely that, for the large values of $|\Phi 
_{max}|$ that are necessary, there will be a proton atmosphere which is never 
exhausted.  In this instance, steady-state plasma formation and low-altitude 
acceleration  appear probable, though subject to possible instabilities at those 
higher altitudes where the coherent radiation is formed.  The same statement can 
be made about all case (i) pulsars whether or not spontaneous CR pair formation  
is supported.  Microstructure with $10^{-4} - 10^{-3}$ s timescales is likely to 
be a consequence of such higher-altitude instabilities.  The reason for this 
assumption is that the case (i) boundary condition on $\Phi$ needs only an 
electron density on the surface separating open from closed magnetospheres.
The small electron cohesive energy means that
there is no reason why this should not be maintained at all times, as is also 
true for the Goldreich-Julian electron current density.  For this boundary 
condition, there is no obvious way in which non-electromagnetic time-constants 
can influence plasma formation and acceleration at altitudes less than $\sim 
10^{6}$ cm above the polar cap. It is possible, perhaps, to assign either this 
boundary condition or case (ii) with spontaneous CR pair formation
to those pulsars in the systematic study of Weltevrede et al (2006), 
approximately one half of the total in number, that do not exhibit subpulse 
drift.

\bsp

\label{lastpage}

\end{document}